\documentclass[doublecol]{epl2}  

\title{Variations of polarisation in external electrostatic fields} \shorttitle{} 

\author{S.Selenu\inst{1}} \shortauthor{S. Selenu}

\institute{                     \inst{1} Dipartimento di Fisica, Universita`
di Cagliari  snail-mail : Cittadella Universitaria, I-09042 Monserrato (CA),
Italy } 

\abstract{It is nowadays a quite diffuse idea that variations of polarisation in condensed matter theory are related to a "Berry phase". The derivation of the latter geometric phase is correct $\it{only~if}$ the restrictive periodic gauge\cite{KS-V} is used for its derivation, aside giving rise to a "quantum of polarisation" that is of difficult theoretical interpretation. Its derivation has not been demonstrated in the general case of an external homogeneous electric field interacting with the electronic field. In the present paper we give a brief derivation of the polarisation differences in a general manner than in\cite{KS-V} allowing for a general boundary conditions freedom showing that is possible to define it even in the presence of an external electric field interacting with the electronic field and writing equations in a way directly implementable in modern first principles codes.}

\begin{document}

\maketitle

\section{Introduction} A revolution of old concepts of classical electrostatics 
has been achieved by the work of several authors \cite{KS-V,{Resta-ferro}},
where we cite in particular that of  Resta\cite{Resta1,Resta2}. There the
concept of $\it{variations}$ of polarisation $\Delta \bf{P}$ as a geometrical quantum phase
has been introduced, and only the definition of the polarisation  of the
electronic field in crystalline materials in absence of external electric
field has been considered. Following a different approach, we shall define
the polarisation differences in a quite general manner allowing for its calculation when it is the case of an external homogeneous electrostatic field interacting with the electronic system. Also, in our derivation we still consider changes of $\Delta \bf{P}$ which occurs upon making a change in the Hamiltonian (Kohn-Sham Hamiltonian if we consider a density functional theory context) of a physical system writing equations in a way directly implementable in modern first principles codes.

\section{Polarisation differences}
As it is already well known\cite{Resta-ferro} the electric dipole of a macroscopic but finite piece of matter is not a bulk property as the latter depends upon truncation and shape of the chosen sample. We agree that only variations of the latter are experimentally accessible where it is supposed that the state of the physical system is specified by a certain parameter $\lambda$. We consider as a formal definition of the electronic dipole the following expression:

\begin{equation}
\label{Macdipole}
{\bf{P}}(\lambda)=\frac{1}{\Omega}\int d\bf{r} \bf{r} \rho_{\lambda}(\bf{r})
\end{equation}

where $\Omega$ is the volume of the system and $\rho_{\lambda}(\bf{r})$ is the total electric charge due to the electronic and ionic contributions. Making use of the Born-Oppenheimer approximation we can divide the electronic and ionic contributions in expression (\ref{Macdipole}) and express the electronic charge contribution as follows: 

\begin{equation}
\label{charge}
\rho_{el}=e \int \frac{d\bf{k}}{(2\pi)^{3}} \sum_{n} f_{n} \Psi^{*}_{n,\bf{k}}\Psi_{n,\bf{k}}
\end{equation}

where $|\Psi_{n,\bf{k}}>$ are the eigenstates of the Hamiltonian (dependent on $\lambda$)and $f_{n}$ are the occupation factors. We always assume eigenstates of the Hamiltonian of being in the Bloch form:

\begin{equation}
\label{eq1} |\Psi_{n,{\bf{k}}}>=e^{i{\bf{k}} \cdot {{\bf{r}}}} |u_{n,{\bf{k}}}>
\end{equation}

Within this  $\it{gauge}$  representation it is naturally apparent the parametric dependence  of electronic
eigenstates on the wave vector $\bf{k}$. Here $\it{quasi~periodic}$\cite{pow} boundary conditions are concerned, where eigenstates $|\Psi_{n,{\bf{k}}}>$ are not periodic while $|u_{n,{\bf{k}}}>$ are periodic on a lattice and orthonormal in the unit cell but $\it{not~periodic}$ in the reciprocal space. Within this  boundary conditions the eigenstates are differentiable and wave vectors $\bf{k}$ are continuous and can be treated at the same foot of the $\bf{r}$ coordinate. On the other hand our derivation of the polarisation differences can equally be brought forward assuming the periodicity of $|u_{n,{\bf{k}}}>$ in reciprocal space. Let us assume the Hamiltonian of the system being of the form:

\begin{equation}
\label{Hamiltoniana}
H(\lambda)=T+V({\bf{E}}_{0},\lambda)
\end{equation} 

\noindent where ${\bf{E}}_{0}$ is an external homogeneous electrostatic field.

\begin{widetext}[t!]
\begin{equation}
\label{P'1}
{\bf{P}^{'}}_{el}= \frac{-ie\hbar}{N\Omega m}\sum_{\bf{k}} \sum_{n} f_{n} \sum_{m\neq n} \frac{<\Psi_{n,{\bf{k}}}|{\bf{p}}|\Psi_{m,{\bf{k}}}><\Psi_{m,{\bf{k}}}|\frac{\partial H}{\partial \lambda}|\Psi_{n,{\bf{k}}}>}{(\epsilon_{n,\bf{k}}-\epsilon_{m,\bf{k}})^{2}} \\+ c.c.
\end{equation}
\end{widetext}

\begin{widetext}[t!]
\begin{equation}
\label{P'2}
{\bf{P}^{'}}_{el}= \frac{-ie\hbar}{N\Omega m} \sum_{{\bf{k}}} \sum_{n} f_{n} \sum_{m\neq n} \frac{<u_{n,{\bf{k}}}|{\bf{p}}+\hbar {\bf{k}}|u_{m,{\bf{k}}}><u_{m,{\bf{k}}}|\frac{\partial H_{\bf{k}}}{\partial \lambda}|u_{n,{\bf{k}}}>}{(\epsilon_{n,\bf{k}}-\epsilon_{m,\bf{k}})^{2}} \\+ c.c.
\end{equation}
\end{widetext}

\begin{widetext}[t!]
\begin{equation}
\label{commutator1}
<u_{n,{\bf{k}}}|i\nabla_{\bf{k}}|u_{m,{\bf{k}}}>=\frac{i\hbar}{m}\frac{<u_{n,{\bf{k}}}|{\bf{p}}+\hbar{\bf{k}}|u_{m,{\bf{k}}}>}{\epsilon_{n,\bf{k}}-\epsilon_{m,\bf{k}}}
\end{equation}
\end{widetext}

\begin{widetext}[t!]
\begin{equation}
\label{P'3}
{\bf{P}^{'}}_{el}= \frac{-e}{N\Omega} \sum_{{\bf{k}}} \sum_{n} f_{n} \sum_{m\neq n} \frac{<u_{n,{\bf{k}}}|i\nabla_{{\bf{k}}}|u_{m,{\bf{k}}}><u_{m,{\bf{k}}}|\frac{\partial H_{\bf{k}}}{\partial \lambda}|u_{n,{\bf{k}}}>}{\epsilon_{n,\bf{k}}-\epsilon_{m,\bf{k}}} \\+ c.c.
\end{equation}
\end{widetext}

\noindent We consider by hypothesis that the system is in a $\it{steady~state}$\cite{Feynman}, and focus our attention only on the electronic part of the electronic dipole trying to calculate the polarisation derivatives with respect to the parameter $\lambda$.
Bearing in mind eq.(\ref{charge}) we can easily determine derivatives\cite{Resta-ferro,KS-V} of the dipole of the system with respect to $\lambda$ in a finite sample and then take the thermodynamic limit. The dipole derivative is reported in eq.(\ref{P'1}), where $\epsilon_{n,\bf{k}}$ are the eigen-energies of the electronic system, c.c. represents the complex congiugate, $\Omega$ the volume of the unit cell, N the number of cells of the system, $\bf{p}$ is the momentum operator. 

Up to know our derivation is similar to Resta derivation\cite{Resta-ferro} apart from a different choice of the boundary conditions for eigenstates $|\Psi_{n,{\bf{k}}}>$ and eigen-amplitudes $|u_{n,{\bf{k}}}>$, and the assumption that the electronic system is in a steady state and interacting with an external homogeneous electrostatic field. 
At this point we can recast eq.(\ref{P'1}) in terms of eigen-amplitudes  $|u_{n,{\bf{k}}}>$ and report its expression in eq.(\ref{P'2}). There $H_{\bf{k}}$ is the effective Hamiltonian $H_{\bf{k}}=e^{-i\bf{k}\cdot{\bf{r}}}He^{i\bf{k}\cdot{\bf{r}}}$.

Before to proceed further we consider the matrix identity already demonstrated in \cite{Zak} and reported in eq.(\ref{commutator1}). Let us now substitute eq.(\ref{commutator1}) in  eq.(\ref{P'2}) and obtain eq.(\ref{P'3}). At this point we can easily estimate the infinite sum in eq.(\ref{P'3}) making use of Salem sum rule\cite{Salem}
and recast eq.(\ref{P'3}) in a more compact and easy calculable form obtaining:

\begin{equation}
\label{P'4}
{\bf{P}^{'}}_{el}= \frac{e}{N\Omega} \sum_{{\bf{k}}} \sum_{n} f_{n} \frac{\partial}{\partial \lambda}
<u_{m,{\bf{k}}}|i\nabla_{{\bf{k}}}|u_{n,{\bf{k}}}>
\end{equation}

\noindent then integrating in $\lambda$ between two different states labelled by $\lambda=0$ and $\lambda=1$ and taking the thermodynamic limit we can recast the integral of eq.(\ref{P'4}) as follows:

\begin{equation}
\label{DPl}
\Delta {\bf{P}}_{el}={\bf{P}}^{(1)}_{el}-{\bf{P}}^{(0)}_{el}
\end{equation}

being,
 
\begin{equation}
\label{DP}
{\bf{P}}_{el}= e \int \frac{d\bf{k}}{(2\pi)^{3}} \sum_{n} f_{n} <u_{m,{\bf{k}}}|i\nabla_{{\bf{k}}}|u_{n,{\bf{k}}}>
\end{equation}

that is formally the same result obtained in \cite{KS-V} for the case of $\it{no}$ electric field interacting with the electronic field. 

Our result it is general and allows for calculations of polarisation differences even in the case of an external electric field interacting with the electronic field. Aside, we do not need anymore ask for the periodic gauge \cite{KS-V}, i.e eigen-amplitudes $|u_{n,{\bf{k}}}>$ periodic in $\bf{k}$. We achieve the result that polarisation differences are $\it{always}$ calculable by eq.(\ref{DP}) allowing us to make use of a more flexible boundary condition. On the other hand our derivation still holds even in the case we use a periodic gauge. In turn our result allows for the elimination of the so called "quantum of polarisation"\cite{KS-V} that is basically due to the very restrictive periodic gauge, attracting attention on the fact that the presence of an external electric field breaks the translational symmetry in the direction of the field and that perhaps the correct boundary condition of the physical problem might not be periodic in reciprocal space.   
   
\section{Summary} During the analysis of the problem started by considering
the interaction of the electronic field with  an  external electric
field it has been shown how to calculate variations of polarisation.  We firstly make use of the expression derived by Resta for the polarisation derivatives then we calculate, in a very general manner without resorting to the restrictive periodic gauge, the polarisation differences. 
Our result is $\it{independent}$ from derivation brought forward in ref\cite{KS-V} where it was the case of no interaction of the electronic field with an external homogeneous electric field. In that case it was considered the polarisation difference as a Berry phase of the system but the latter derivation needs of a very restrictive boundary condition, the so called periodic gauge. Without the above mentioned boundary condition the algorithm proposed in ref\cite{KS-V} $\it{fails}$ while our derivation still holds allowing for a general gauge freedom and also allowing for the elimination of the concept of "quantum of polarisation", the latter being  theoretically difficult to be interpreted. Aside, we wrote equations in a way directly implementable in modern first principles codes.

\acknowledgments I would like to acknowledge the Atomistic Simulation Centre
at Queen's University Belfast for hospitality during the course of this work,
and I am grateful to Prof. M.W.Finnis for helpful discussions. This work has
been supported by the ESF.


\begin{thebibliography}{0} \expandafter\ifx\csname
natexlab\endcsname\relax\def\natexlab#1{#1}\fi \expandafter\ifx\csname
bibnamefont\endcsname\relax \def\bibnamefont#1{#1}\fi \expandafter\ifx\csname
bibfnamefont\endcsname\relax \def\bibfnamefont#1{#1}\fi
\expandafter\ifx\csname citenamefont\endcsname\relax
\def\citenamefont#1{#1}\fi \expandafter\ifx\csname url\endcsname\relax
\def\url#1{\texttt{#1}}\fi \expandafter\ifx\csname
urlprefix\endcsname\relax\def\urlprefix{URL }\fi
\providecommand{\bibinfo}[2]{#2} \providecommand{\eprint}[2][]{\url{#2}}





\bibitem{KS-V} \bibinfo{author}{\bibfnamefont{R.D.}~\bibnamefont{King-Smith}},
\bibinfo{author}{\bibfnamefont{D.}~\bibnamefont{Vanderbilt}},
\emph{\bibinfo{title}{Theory of polarization of crystalline solids}},
\bibinfo{journal}{Phys. Rev. B} \textbf{\bibinfo{volume}{47}},
\bibinfo{pages}{R1651} (\bibinfo{year}{1993}).

\bibitem{Resta-ferro} \bibinfo{author}{\bibfnamefont{R.}~\bibnamefont{Resta}},
\emph{\bibinfo{title}{Theory of the electric polarization in crystals}},
\bibinfo{journal}{Ferroelectrics} \textbf{\bibinfo{volume}{136}},
\bibinfo{pages}{51} (\bibinfo{year}{1992}).


\bibitem{Resta1} \bibinfo{author}{\bibfnamefont{R.} \bibnamefont{Resta}}
\emph{\bibinfo{title}{Berry Phase in Electronic Wavefunctions}},
(\bibinfo{publisher}{Troisieme cycle Lecture Notes (Ecole Politechnique
Federale, Lausanne, Switzerland, 1996). Avaiable on line (194K) at the URL:
http://ale2ts.ts.infn.it:6163/~resta/publ/notes.trois.ps.gz}.

\bibitem{Resta2} \bibinfo{author}{\bibfnamefont{R.}~\bibnamefont{Resta}},
\emph{\bibinfo{title}{Macroscopic polarization in crystalline dielectrics: the
geometric phase approach}},  \bibinfo{journal}{Rev. of Mod. Phys.}
\textbf{\bibinfo{volume}{66}}, \bibinfo{pages}{899} (\bibinfo{year}{1994}).

\bibitem{pow} \bibinfo{author}{\bibfnamefont{S.}~\bibnamefont{Selenu}},
\emph{\bibinfo{title}{Quantum electro-mechanics: the power density}},  \bibinfo{journal}{Europhys. Lett.}
\textbf{\bibinfo{volume}{to be published}}, \bibinfo{pages}{} (\bibinfo{year}{2008}).

\bibitem{Feynman} \bibinfo{author}{\bibfnamefont{R.P.}~\bibnamefont{Feynman}},
\emph{\bibinfo{title}{Forces in Molecules}}, \bibinfo{journal}{Phys. Rev.}
\textbf{\bibinfo{volume}{56}}, \bibinfo{pages}{340} (\bibinfo{year}{1939}).

\bibitem{Zak} \bibinfo{author}{\bibfnamefont{J.}~\bibnamefont{Zak}},
\emph{\bibinfo{title}{Berry's phase in the effective-Hamiltonian theory of solids}},
\bibinfo{journal}{Phys. Rev. B} \textbf{\bibinfo{volume}{40}},
\bibinfo{pages}{3156} (\bibinfo{year}{1989}).


\bibitem{Salem} \bibinfo{author}{\bibfnamefont{L.}~\bibnamefont{Salem}},
\emph{\bibinfo{title}{Quantum-Mechanical sum rule for infinite sums involving the Operator $\frac{\partial H}{\partial \lambda}$}},\bibinfo{journal}{Phys. Rev.} \textbf{\bibinfo{volume}{125}},
\bibinfo{pages}{1788} (\bibinfo{year}{1962}).




\end{thebibliography}
\end{document}